\documentclass[aps,twocolumn,showpacs,superscriptaddress,preprintnumbers]{revtex4}
\usepackage{amssymb}
\usepackage{amsmath}
\usepackage{graphicx}
\usepackage{epsfig}

\begin{document}

\title{Effective boson-spin model for nuclei ensemble based universal
quantum memory}
\author{Zhi Song$^{1,a}$, Peng Zhang$^{2}$, Tao Shi$^{1}$ and Chang-Pu Sun$%
^{1,2,b,c}$}

\begin{abstract}
We study the collective excitation of a macroscopic ensemble of polarized
nuclei fixed in a quantum dot. Under the approximately homogeneous condition
that we explicitly present in this paper, this many-particle system behaves
as a single mode boson interacting with the spin of a single conduction band
electron confined in this quantum dot. Within this effective spin-boson
system, the quantum information carried by the electronic spin can be
coherently transferred into the collective bosonic mode of excitation in the
ensemble of nuclei. In this sense, the collective bosonic excitation can
serve as a stable quantum memory to store the quantum spin information of
electron.
\end{abstract}

\pacs{PACS number: 73.21.La,03.65.-w, 03.67.¨Ca, 76.70.¨Cr }
\maketitle

\address{ $^{1}$Department of Physics, Nankai University, Tianjin
300071, China\\
$^{2}$ Institute of Theoretical Physics, The Chinese Academy of
Science, Beijing, 100080, China}

\section{Introduction}

In the current development of quantum information science and
technologies, people have devoted much effort searching for the
optimal system serving as a long-lived quantum memory to store the
quantum information carried by a quantum system with short
decoherence time \cite{q-infor}. A universal quantum information
storage can be understood as a physical process to encode the
states of each qubit (rather than the general quantum state) into
the states of the quantum memory with much longer decoherence time
than the life time of qubit \cite{za-store}; or transform the
quantum information carried by a quantum system (such as photons)
which is difficult to manipulate to an easily controllable system
(such as the localized atomic ensemble) \cite{sun-prl}. Such
quantum information storages are absolutely
necessary in both measurement based quantum computation schemes \cite%
{Knill,zhou} and two-qubit gate-based computation schemes \cite{U-gate}.

In the past years the collective excitation of the ensemble of atoms have
been proposed to serve as quantum memory for photon information \cite%
{Lukin-RMP}. Several experiments \cite{lui,lukin-exp,kimble} have
already demonstrated the central principle of this scheme. These
schemes work to record the Fock states of photon or their coherent
superpositions. In this paper we will pay attention to the
universal quantum storage (called qubit storage) that stores the
basic two-level state, the state of qubit rather than a general
quantum state \cite{za-store}. The universality of the qubit
storage lies on the fact that a general quantum state can be
encoded as the state of multi-qubits and the corresponding quantum
logic operations can be decomposed into the "quantum networks"
which are the product of the fundamental operations defined with
respect to the qubits \cite{U-gate}.

In usual, the foundation of a universal scheme of quantum information
storage depends on whether one can discover a new quantum system with much
long decoherence time as the universal quantum memory. Recently a novel
protocol for universal quantum information storage was presented based on
the nanomechanical resonator interacting with charge qubits. As the
universal quantum memory, the nanomechanical resonator behaves as a single
mode harmonic oscillator and its coupling to charge qubit is just described
by the Jaynes-Cummings (JC) model \cite{cleland}. Such spin-boson
interaction forms the basis for ion-trap based computation schemes as well
\cite{ion}. These idealized schemes motivate us to seek another more
practical protocol based on collective bosonic excitation in various
physical systems. We notice that a mesoscopic system that consists of finite
nuclear spins attached in a quantum dot has been proposed to realize a
long-lived quantum memory in this universal way \cite%
{uqm-lukin,uqm-zoller,uqm-exp}. The present article will start from this
basic idea and then work on the macroscopic limit that the number of
polarized nuclei is very large so as to be treated approximately as infinite.

We will show that, under two independent sufficiently approximately
homogeneous conditions, the collective excitation of a macroscopic ensemble
of polarized nuclei fixed in a quantum dot can behave as a single bosonic
mode. In this sense, confined in this quantum dot, the spin of a single
conduction band electron interacts with this collective excitation and then
forms an effective spin-boson system. It demonstrates a dynamic process to
coherently store the quantum information carried by the electronic spin in
the collective bosonic mode of the nuclei ensemble. Then the collective
excitation of the nuclei ensemble can serve as a universal quantum memory to
store the quantum information of spin state of electron.

\section{Boson Realization of Collective Excitation in the Ensemble of
Polarized Nuclei}

We can consider the ensembles of $N\sim10^{3-5}$ polarized nuclei with spin $%
I_{0}$, which are fixed in a charged quantum dot and interact with a single
conduction band electron confined in this dot (Fig. 1). There exists a
hyperfine contact interaction between the s-state conduction electron and
the fixed nuclei. When a static magnetic field is applied to the dot, the
effective Hamiltonian for the total system reads

\begin{eqnarray}
H &=&\Omega _{z}\sigma _{z}+\omega _{z}\sum_{j=1}^{N}I_{z}^{(j)}  \notag \\
&&+\sigma _{z}\sum_{j=1}^{N}g_{j}I_{z}^{(j)}+\sigma _{+}\sum_{j=1}^{N}\frac{%
g_{j}}{2}I_{-}^{(j)}+h.c,
\end{eqnarray}%
where the operators
\begin{equation}
\mathbf{I}^{(i)}=(I_{x}^{(i)},I_{y}^{(i)},I_{z}^{(i)})
\end{equation}%
and
\begin{equation}
\mathbf{\sigma }=(\sigma _{x},\sigma _{y},\sigma _{z})
\end{equation}%
describe the spins of the nucleus at $i$th site and the conduction electron
respectively. The coefficients $\Omega _{z}$ and $\omega _{z}$ are the
Larmor precession frequencies of nucleus and electron which are linearly
determined by the applied external magnetic field. The strength $g_{i}$ of
the hyperfine interaction depends on the local value of the norm $\left\vert
\psi (x_{i})\right\vert ^{2}$ at the position $x_{i}$ of $i$th nucleus,
while $\psi (x_{i})$ is the wave packet of a single electron inside the dot.
In this article, we take the average of couplings as $\overline{g}%
=\sum_{i}g_{i}/N=A/N$ . Here, we have also used the spin-flip
operators
\begin{equation}
\sigma _{\pm }=\left( \sigma _{x}\pm i\sigma _{y}\right) /2
\end{equation}%
and%
\begin{equation}
I_{\pm }^{\left( j\right) }=\left( I_{x}^{\left( j\right) }\pm
iI_{y}^{\left( j\right) }\right) /2.
\end{equation}%
In the following discussion, the eigenstates of $\sigma _{z}$ and $%
I_{z}^{\left( j\right) }$ are denoted as $\left\vert \uparrow \right\rangle
_{e}$, $\left\vert \downarrow \right\rangle _{e}$ and $\left\vert
m\right\rangle _{j}$ which satisfy
\begin{equation}
\sigma _{z}\left\vert \uparrow \right\rangle _{e}=\left\vert \uparrow
\right\rangle _{e},\sigma _{z}\left\vert \downarrow \right\rangle
_{e}=-\left\vert \downarrow \right\rangle _{e}
\end{equation}%
and
\begin{equation}
I_{z}^{\left( j\right) }\left\vert m\right\rangle _{j}=m\left\vert
m\right\rangle _{j},m=I_{0},I_{0}-1,...,-I_{0}.
\end{equation}%
\vspace{0.5cm}

\begin{figure}[tbp]
\hspace{24pt}\includegraphics[width=6cm,height=3cm]{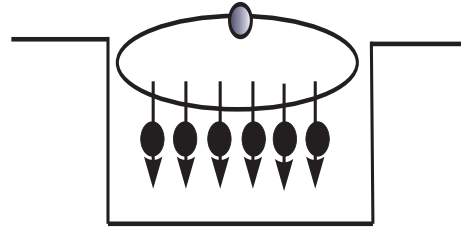} \vspace*{0.5cm}
\caption{The polarized nuclei interacting with an electron in a quantum dot.
Because of the overlap between the electronic wave function and the nuclear
wave functions, an effective spin-spin coupling between the electron and the
nuclei is induced.}
\end{figure}

An obvious observation seen from the above expression of the Hamiltonian (1)
is that all nuclei wholly couple to a single electron spin. Then we can
introduce a pair of collective operators

\begin{equation}
B=\frac{\sum_{i=1}^{N}g_{i}I_{-}^{(i)}}{\sqrt{2I_{0}\sum_{j}g_{j}^{2}}}
\end{equation}%
and its conjugate $B^{+}$ to depict the collective excitations in the
ensemble of nuclei with spin $I_{0}$ from its polarized initial state%
\begin{equation}
\left\vert G\right\rangle =\left\vert M=-NI_{0}\right\rangle
=\prod\limits_{i=1}^{N}\left\vert -I_{0}\right\rangle _{i},
\end{equation}%
where $M$ is the eigen value of the $z$-component of total nuclei spin $%
I_{z}=\sum_{i=1}^{N}I_{z}^{(j)}$, which denotes the saturated
ferromagnetic state of nuclei ensemble.

Now we can show that the collective excitations depicted by $B$
and $B^{+}$ can behave as bosons under the "quasi-homogeneity"
conditions in the low excitation limit (we will explicitly present
this as follows). In fact, in the previous investigations
\cite{sun-prl,sun-pra,sun-prb}, we have proved that, if the
coupling is homogeneous or in a periodic way, the collective
operators $B$ and $B^{\dagger}$ can indeed be considered as boson
operators in the low excitation and macroscopic limit
$n/N\rightarrow 0$, where $n$ is the number of excitation from
ground state $\left\vert G\right\rangle$. The number $n$
characterizes the number of collective excitations of the nuclei
ensemble, which is defined through the eigenvalue
$m_{n}=-NI_{0}+1/2+n$ of the $z$-component of total spin
\begin{equation}
J_{z}=\sigma _{z}+I_{z},
\end{equation}%
where $n=0,1,2,\cdots ,2NI_{0}-1.$ It is obvious that $J_{z}$ is a
"good quantum number" for the Hamiltonian (1) since $\left[
J_{z},H\right] =0$. Here, we do not include the saturated
ferromagnetic states with $J_{z}=\pm (NI_{0}+1/2)$. It is noticed
that $n$ is actually the number of excitations in the system,
regardless of mode $B$ or others. There is an intuitive argument
that if the $g_{i}$s have different values, but the distribution
is "quasi-homogeneous", $B$ and $B^{\dagger }$ can also be
considered as boson operators satisfying

\begin{equation}
\lbrack B,B^{+}]\rightarrow 1  \label{aa}
\end{equation}%
approximately. In the following discussion, we will provide two descriptions
\ for the "quasi-homogeneity" condition, under which Eq. (\ref{aa}) holds in
the limit $n/N\rightarrow 0$.

To this end we re-write the commutator of $B$ and $B^{\dagger }$ as%
\begin{equation}
\lbrack B,B^{+}]=-\frac{\sum_{j}g_{j}^{2}I_{z}^{\left( j\right) }}{%
I_{0}\sum_{j}g_{j}^{2}}\equiv 1-F(N,n),
\end{equation}%
where
\begin{equation}
F(N,n)=\frac{\sum_{j}g_{j}^{2}\left[ I_{z}^{\left( j\right) }+I_{0}\right] }{%
I_{0}\sum_{j}g_{j}^{2}}.
\end{equation}%
Since $g_{j}^{2}\leq g_{\max }^{2}$, $F=F(N,n)$ can be estimated
as
\begin{equation}
F\leq \frac{g_{\max }^{2}\left( I_{z}+NI_{0}\right) }{\overline{g^{2}}I_{0}N}%
\leq \frac{g_{\max }^{2}n}{\overline{g^{2}}I_{0}N},
\end{equation}%
where $\overline{g^{2}}=\sum_{j}g_{j}^{2}/N$ is the average over set $%
\{g_{j}^{2}\}$. Here we have used the definition of $n$: $I_{z}+I_{0}N=n$ or
$n+1$ with respect to the electronic spin up or spin down and the condition $%
NI_{0}>>1$. Therefore, it is easy to see that when $g_{\max }^{2}/\overline{%
g^{2}}\sim 1$ in the limit $n/N\rightarrow 0,$ we have $F(N,n)\rightarrow 0$
and then $[B,B^{+}]\simeq 1$. Based on the above argument, the first
"quasi-homogeneity" condition can be obtained as%
\begin{equation}
\frac{g_{\max }^{2}}{\overline{g^{2}}}\quad \sim 1.  \label{bb}
\end{equation}%
We would like to point out that the above condition corresponds to the
physically relevant case of a quantum dot since $g_{i}$ is proportional to
the norm $\left\vert \psi (x_{i})\right\vert ^{2}$ at the position of nuclei
\ for the wave packet $\psi (x_{i})$ of a quasi-free electron moving inside
the dot.

However, the above "quasi-homogeneity" condition is not necessary
and we can find another independent one as follows. By a
straightforward calculation we can also re-express $F(N,n)$ as
\begin{equation}
F=1-\frac{\left( I_{z}+I_{0}N\right) }{I_{0}N}-\frac{\sum_{j}I_{z}^{\left(
j\right) }\left[ g_{j}^{2}-\overline{g^{2}}\right] }{I_{0}N\overline{g^{2}}}.
\label{a}
\end{equation}%
Since the term $\left( I_{z}+I_{0}N\right) /\left(
I_{0}N\right)\sim n/\left( I_{0}N\right)$ for $N>>1$ and
$\left\vert \left\langle I_{z}^{\left( j\right) }\right\rangle
\right\vert \leq I_{0}$, the upper limit of the second term in the
right side of Eq. (\ref{a}) can be estimated as
\begin{equation}
\left\vert \frac{\sum_{j}I_{z}^{\left( j\right) }(g_{j}^{2}-\overline{g^{2}})%
}{I_{0}N\overline{g^{2}}}\right\vert \leq \left( \overline{\delta g^{2}}%
\right) /\overline{g^{2}}
\end{equation}%
in terms of the absolute value deviation
\begin{equation}
\overline{\delta g^{2}}=\frac{1}{N}\sum_{j}\left\vert g_{j}^{2}-\overline{%
g^{2}}\right\vert
\end{equation}%
of $g_{j}^{2}$. Therefore it is obvious that $[B,B^{+}]\rightarrow
1$ in the low excitation limit $n/N\rightarrow 0$ when another
"quasi-homogeneity"
condition%
\begin{equation}
\frac{\overline{\delta g^{2}}}{\overline{g^{2}}}\rightarrow 0  \label{dd}
\end{equation}%
holds.

It is pointed out that, both of the two "quasi-homogeneity" conditions (\ref%
{bb}) and (\ref{dd}) are sufficient conditions for the boson commutation
relation $[B,B^{+}]=1$ , but they are independent with each other and we can
obtain neither of them from another. There are some cases in which one of
the two conditions is satisfied, but another is violated. For instance, in
the case with $N\sim 10^{4}$, if $g_{1}=g_{2}=...=g_{N-1}=g$ and $g_{N}=10g$
then we have $\overline{g^{2}}=1.01g^{2}$, $g_{\max }^{2}=100g^{2}$ and $%
\overline{\delta g^{2}}\approx 0.02g^{2}$. It is apparent that the condition
(\ref{bb}) is violated, but the condition (\ref{dd}) is satisfied since $%
g_{\max }^{2}/\overline{g^{2}}\approx $ $100$ and $\overline{\delta g^{2}}/%
\overline{g^{2}}\approx 0.02$. In another example , we take $%
g_{1}=g_{2}=...=g_{N/2}=g$ and $g_{N/2+1}=g_{N/2+2}=...=g_{N}=3g$, then we have $%
\overline{g^{2}}=5g^{2}$, $g_{\max }^{2}=9g^{2}$ and $\overline{\delta g^{2}}%
\approx 4g^{2}$. This is a physically relevant case when the size of
electron wave function fixed and the nuclear spin density is increased. It
indicates that the condition (\ref{dd}) is violated, but the condition (\ref%
{bb}) is satisfied in this case.

\section{Effective Hamiltonian Decorated by External Magnetic Field}

As mentioned above the $z$-component of total spin $J_{z}=\sigma _{z}+I_{z}$
is conserved and thus we can classify the total Hilbert space for the
ensemble of polarized nuclei according to the excitation number $n.$ In the
following we denote the eigenspace of $J_{z}$ with eigenvalue $m_{n}$ by $%
V_{n}$. Then $V_{n}$ can be decomposed into a direct sum of two eigen-spaces
\begin{equation}
V_{n+}=Span\{\left\vert g_{k}^{\left( n\right) }\right\rangle |k=1,2..,\}
\end{equation}
where the basis vectors
\begin{equation}
\left\vert g_{k}^{\left( n\right) }\right\rangle \in \{\left\vert \uparrow
\right\rangle \otimes \left\vert l_{1}l_{2}...l_{N}\right\rangle
|\sum_{j}l_{j}=-NI_{0}+n\}
\end{equation}%
and%
\begin{equation}
V_{n-}=Span\{\left\vert f_{k}^{\left( n\right) }\right\rangle |k=1,2..,\}
\end{equation}
where the basis vectors

\begin{equation}
\left\vert f_{k}^{\left( n\right) }\right\rangle \in \{\left\vert \downarrow
\right\rangle \otimes \left\vert l_{1}l_{2}...l_{N}\right\rangle
|\sum_{j}l_{j}=-NI_{0}+n+1\}
\end{equation}%
of $I_{z}$, i.e., $V_{n}=V_{n+}\oplus V_{n-}$. Then the
Hamiltonian (1) can
be decomposed into three parts in the invariant subspace $V_{n},$ namely,%
\begin{equation}
H=H_{R}+H_{S}+H_{p}.
\end{equation}%
Each part $H_{R},H_{S}$ and $H_{p}$ can be described as follows: The first
part
\begin{equation}
H_{R}=\Omega \left( \sigma _{+}B+\sigma _{-}B^{+}\right)
\end{equation}%
is a resonate JC Hamiltonian with the collective Rabi frequency
\begin{equation}
\Omega =\sqrt{I_{0}\sum_{j}\frac{g_{j}^{2}}{2}}
\end{equation}
coupling the electron spin to the collective excitation.
Associated with the non-collective excited states $\left\vert
g_{k}^{\left( n\right) }\right\rangle$, $\left\vert f_{k}^{\left(
n\right) }\right\rangle $ and the corresponding composite energies

\begin{eqnarray}
\mu _{g}(n) &=&\frac{\Omega _{z}}{2}+\omega _{z}(m_{n}-\frac{1}{2})-\frac{N%
\overline{g}I_{0}}{2},  \notag \\
\mu _{f}(n) &=&-\frac{\Omega _{z}}{2}+\omega _{z}(m_{n}+\frac{1}{2})+\frac{N%
\overline{g}I_{0}}{2},
\end{eqnarray}%
the second part

\begin{eqnarray}
H_{S} &=&\Omega _{z}\sigma _{z}+\omega _{z}\sum_{j=1}^{N}I_{z}^{(j)}-\sigma
_{z}\sum_{j=1}^{N}g_{j}I_{0} \\
&=&\mu _{g}\left( n\right) \sum_{k}\left\vert g_{k}^{\left( n\right)
}\right\rangle \left\langle g_{k}^{\left( n\right) }\right\vert +\mu
_{f}\left( n\right) \sum_{k}\left\vert f_{k}^{\left( n\right) }\right\rangle
\left\langle f_{k}^{\left( n\right) }\right\vert  \notag
\end{eqnarray}%
is derived from the first and second terms of the original Hamiltonian and
also operates within the subspace $V_{n}.$ In the third part
\begin{eqnarray}
H_{p} &=&\sigma _{z}\sum_{j=1}^{N}g_{j}(I_{z}^{(j)}+I_{0})  \label{b} \\
&=&\sum_{k}\sum_{j=1}^{N}\frac{g_{j}}{2}\left( M_{kn}^{\left( j\right)
}+I_{0}\right) \left\vert g_{k}^{\left( n\right) }\right\rangle \left\langle
g_{k}^{\left( n\right) }\right\vert  \notag \\
&&-\sum_{k}\sum_{j=1}^{N}\frac{g_{j}}{2}\left( M_{kn}^{\prime \left(
j\right) }+I_{0}\right) \left\vert f_{k}^{\left( n\right) }\right\rangle
\left\langle f_{k}^{\left( n\right) }\right\vert  \notag
\end{eqnarray}%
$M_{kn}^{\left( j\right) }$ ($M_{kn}^{\prime \left( j\right) }$) is the $c$%
-number which describes the $z$ component of the $j$-th nuclear spin in the
state $\left\vert g_{k}^{\left( n\right) }\right\rangle $ ( $\left\vert
f_{k}^{\left( n\right) }\right\rangle $).

We observe that the interaction part $H_{JC}=H_{R}+H_{S}$ is very similar to
the Jaynes-Cummings (JC) Hamiltonian in cavity QED describing the
interaction between the two-level atom and single-mode electromagnetic field
in the rotating wave approximation. In order to create the entanglement
between electron spin and collective bosons, one can adjust the external
field $B_{0}$ so that $\mu _{f}(n)=\mu _{g}(n)$, that is

\begin{equation}
\Omega _{z}=\omega _{z}+N\overline{g}I_{0}
\end{equation}
In this case $H_{JC}$ $=\oplus _{n}H_{JC}^{[n]\text{ }}$on the whole space $%
V=\oplus _{n}$ $V_{n}$ can be reduced to the irreducible parts

\begin{equation}
H_{JC}^{[n]}\sim H_{R}+n\omega _{z}  \label{aaaa}
\end{equation}%
for $\omega _{z}=\Omega _{z}-N\overline{g}I_{0}$. Correspondingly the
dynamics of the total system can also be constrained within the invariant
subspace $V_{n}$ and then the last term $n\omega _{z}$ independent of $%
\sigma _{+}$and $B$ can be ignored since it can not contribute to the
dynamics of the system significantly.

To consider the effectiveness of the above qubit storage protocol, we need
to analyze the role of other collective modes orthogonal to the basic
collective model defined by $B$ and $B^{+}$. These auxiliary collective
modes complement the "$B$" mode to generate a complete Hilbert space of the
nuclei ensemble.In fact, we can generally construct the complete set of
creation and annihilation operators $C_{k}^{+}$ and $C_{k}$ ( $k=1,2,...,N$
) including $C_{0}=B$ and all auxiliary modes as
\begin{equation}
C_{k}=\frac{\sum_{i=1}^{N}h_{i}^{[k]}I_{+}^{(i)}}{\sqrt{2I_{0}%
\sum_{j}h_{j}^{2}}}.
\end{equation}%
where
\begin{equation}
\mathbf{h}^{[k]}=(h_{1}^{[k]},h_{2}^{[k]},...,h_{N}^{[k]})
\end{equation}%
(for $k=1,2,...,N$) are $N$ orthogonal vectors in the $N$-dimension space $%
\mathbf{R}^{N}$, which can be systematically constructed by making
use of the Gramm-Schmidt orthogonalization method starting from
\begin{equation}
\mathbf{h}^{[1]}=(g_{1},g_{2,}...,g_{N})\in \mathbf{R}^{N}.
\end{equation}%
Since $\mathbf{h}^{[k]}\cdot \mathbf{h}^{[j]}=\delta _{kj}$ and the
Gramm-Schmidt orthogonalization can also result in the quasi-homogeneity
conditions
\begin{equation}
\frac{\overline{\delta h^{^{[k]}2}}}{\overline{h^{^{[k]}2}}}\sim 0,or\qquad
\frac{h_{\max }^{^{[k]}2}}{\overline{h^{^{[k]}2}}}\sim 1\qquad
\end{equation}%
we have
\begin{equation}
\lbrack C_{k},C_{j}^{+}]\rightarrow \delta _{kj}
\end{equation}%
in the large $N$ limit.

For example, one can construct a new boson mode with respect to the existing
mode by the collective excitation $B$ by choosing a new distribution of
coupling constants
\begin{equation}
\{h_{i}|h_{i}=g_{N-i,\text{ }}h_{N-i}=-g_{i},\forall i<\frac{N}{2}\}
\end{equation}%
as a permutation of $\{g_{i}\}$ and then define a independent
boson mode by the collective operator $C$ as Eq. (32). We can
check that both the orthogonal relation
$\sum_{i=1}^{N}g_{i}h_{i}=0$ and the quasi-homogeneity conditions
$\overline{\delta h^{2}}/\overline{h^{2}}\sim 0$ or $h_{\max
}^{2}/\overline{h^{2}}\sim 1$ can be satisfied obviously. Then one
can prove that in each invariant subspace $V_{n}$ with $n\ll N$ ,
there are the typical boson commutation relations
\begin{equation}
\lbrack C,C^{+}]=1,[C,B^{+}]=0.
\end{equation}

Apparently, from the above generalized the Gramm-Schmidt orthogonalization,
the auxiliary boson operators can be expressed as the linear combination of
the spin operators through a matrix transformation $\mathbf{C}=U\mathbf{I}$
for
\begin{eqnarray}
\mathbf{C} &=&\left( B,C_{2},...,C_{N}\right) ^{T}, \\
\mathbf{I} &=&\left( I_{-}^{\left( 1\right) },I_{-}^{\left(
2\right) },...,I_{-}^{\left( N\right) }\right) ^{T},  \notag
\end{eqnarray}%
where $U$ is a unitary (or orthogonal) matrix. Since $\mathbf{C}^{+}\mathbf{%
C=I}^{+}\mathbf{I=}\sum_{j=1}^{N}I_{+}^{\left( j\right)}\mathbf{\ }%
I_{-}^{\left( j\right) }$ under such transformation one can prove
that there exists a constraint
\begin{equation}
B^{+}B+\sum_{k}C_{k}^{+}C_{k}\approx I_{z}+I_{0}N.
\end{equation}%
when they work on the subspace $V_{n}$ with $n\ll N.$%
\begin{equation}
\Omega _{z}=\omega _{z}+N\overline{g}I_{0}.
\end{equation}

Formally, the Hamiltonian in Eq. (\ref{aaaa}) can be rewritten in the whole
space as%
\begin{eqnarray}
H_{JC} &=&H_{R}+\omega _{z}I_{z}+\Omega _{z}\sigma _{z}-\sigma
_{z}\sum_{j=1}^{N}g_{j}I_{0}  \notag \\
&\sim &H_{R}+\omega _{z}\sum_{k}C_{k}^{+}C_{k}+\omega _{z}(B^{+}B+\sigma
_{z})
\end{eqnarray}%
according to the above constraint. The above argument implies no coupling
between the basic mode $B$ and the auxiliary modes $C_{k}$ $($ $%
k=1,2,...,N-1)$ and thus the dependence on $C_{k}$ is trivial in the above
equation. However, all mode coupling terms between the auxiliary modes $%
C_{k} $ and the electron spin qubit occur only in $H_{p}$, which can cause
decoherence of the qubit\cite{loss}. In section VI we will explore this
decoherence mechanism in details.

\section{Validity of the Single Mode Approximation}

First, let's assume that $B$ mode is independent of those auxiliary
collective modes $C_{k}$. To formally diagonalize $H_{JC}$ by
straightforward calculations, one can obtain the eigenvalues of $H_{JC}$
\begin{equation}
E_{\pm }({m_{i}},m)=(m+\sum_{i}m_{i})\omega _{z}\pm \sqrt{I_{0}\left(
m+1\right) \sum_{j}\frac{g_{j}^{2}}{2}}
\end{equation}%
and the corresponding eigenstates:%
\begin{eqnarray}
\left\vert \psi ^{(\pm )}\left( \left\{ m_{i}\right\} ,m\right)
\right\rangle &=&\frac{1}{\sqrt{2m!m_{i}!}}\left( \left\vert \uparrow
\right\rangle _{e}\pm \left\vert \downarrow \right\rangle _{e}\frac{B^{+}}{%
\sqrt{m+1}}\right)  \notag \\
&&\otimes (B^{+})^{m}(C_{i}^{+})^{m_{i}}\left\vert G\right\rangle
\end{eqnarray}%
where we have $\sum_{i}m_{i}+m=n$.

It is also pointed out that, only the "excited" nuclear spins
whose $z$ component values are not $-I_{0}$ have contributions to
the summation in the definition of $H_{p}$ (\ref{b}). Therefore,
because of the low exciton condition, there is an intuitive
argument to show that $H_{p}$ is only a perturbation term. In the
following, this guess can be proved explicitly. Under the
quasi-homogeneity conditions (\ref{bb}) and (\ref{dd}), one has
\begin{equation}
H_{p}\sim \frac{\bar{g}}{2}\left( I_{z}+NI_{0}\right) \sigma _{z},
\end{equation}%
where $\bar{g}=A/N$ is the average coupling strength between the electron
and nuclei. Then the first order energy correction for $H_{p}$ can be
estimated with perturbation theory:%
\begin{equation}
\delta E\left( n\right) \sim \left\langle \psi _{n}^{(\pm )}\right\vert
H_{p}\left\vert \psi _{n}^{(\pm )}\right\rangle \sim \frac{n}{4}\bar{g}.
\end{equation}%
On the other hand, the energy gap between $E_{\pm }(n)$ is
\begin{eqnarray}
\Delta E(\left\{ m_{i}\right\} ,m) &=&\sqrt{I_{0}(m+1)\sum_{j}\frac{g_{j}^{2}%
}{2}}  \notag \\
&\sim &\frac{\overline{g}}{\sqrt{2}}\sqrt{NI_{0}\left( m+1\right) }
\end{eqnarray}%
where we have used the relation $\overline{g^{2}}\sim \bar{g}^{2}$, which
meets the conditions (\ref{bb}) and (\ref{dd}). Therefore, the magnitude of
the contribution of $H_{p}$ can be described by the ratio%
\begin{equation}
\left\vert \frac{\delta E\left( n\right) }{\Delta E(\left\{ m_{i}\right\} ,m)%
}\right\vert \sim \sqrt{\frac{n}{N}}.
\end{equation}%
The above estimation implies that $H_{p}$ can indeed be regarded as
perturbation to $H_{JC}$ in the low excitation case and macroscopic limit $%
n\ll N$. Therefore, we can take $H_{JC}$ as the effective
Hamiltonian of the total system. It is noticed that, there is no
coupling between the mode $B$ and the auxiliary modes $C_{k}$.
Hence we can write down the effective
Hamiltonian of the electron spin and the $B$ mode as%
\begin{equation}
H_{c}=\Omega (\sigma _{+}B+\sigma _{-}B^{+})+\omega _{z}\left[ B^{+}B+\sigma
_{z}-1/2\right] .  \label{ccccc}
\end{equation}

In order to quantitatively evaluate the extent of approximation of the
single mode boson approach and obtain the effective Hamiltonian, the
numerical method is employed to verify our approximate analytical result. We
compute the eigenstates of the original spin-exchange Hamiltonian
\begin{equation}
H_{s}=\frac{1}{2}\sum_{j}g_{j}(\sigma _{+}I_{-}^{(j)}+\sigma _{-}I_{+}^{(j)})
\end{equation}%
and
\begin{equation}
H_{c}-n\omega _{z}=\Omega (\sigma _{+}B+\sigma _{-}B^{+}).
\end{equation}%
for finite $N$ system. Without loss of generality we take a Gaussian type
distribution which satisfies the conditions (\ref{bb}) and (\ref{dd}). In
Fig.2, the spectrums of $H_{s}$ and $H_{c}-n\omega _{z}$ for $N=80$ and $%
I_{0}=1/2$ system in the subspace $J_{z}=-N/2+1$ are plotted in
(a) and (b). It shows that the spectrums are in agreement with
each other. By comparing the numerically exact results with the
analytically approximate ones for the effective spin-boson system,
the numerical result shows that in low excitation and macroscopic
limit with the quasi-homogeneity condition, the single mode boson
effective Hamiltonian (\ref{ccccc}) can work well in describing
the collective excitation of the nuclei ensemble stimulated by the
conduction band electron. We can also understand the difference of
the spectrum structures in Fig. 2 in terms of the concept of
"hardcore boson"
\cite{fisher}. We imagine a model Hamiltonian $H_{b}=H_{s}(I_{-}^{(j)}%
\rightarrow b^{(j)})$, which is obtained by replacing $I_{-}^{(j)}$ ($%
I_{+}^{(j)}$) in Eq. (50) with a set operators
$b^{(j)}(b^{(j)+})$. If they
satisfy the usual commutation relation of bosons that the operators $%
b^{(j)+}(b^{(j)})$ and $b^{(k)}(b^{(k)+})$ on different sites commute with
each other, then $B$ and $B^{+}$ automatically satisfy the boson commutation
relation and then result in the regular spectrum same as to that illustrated
in Fig. 2(b). However, if the bosons are of hardcore, i.e., they are
repulsive with each other at a same site, one can describe them with
vanishing anti-commutators $\{b^{(j)+},b^{(j)+}\}=0$ for the same site and
the vanishing commutators $[b^{(j)\pm },b^{(k)\pm }]=0$ for different sites.
In this case the repulsive interaction with hardcore feature will widen the
original spectral lines to form the similar band structure in the energy
spectrum.

\begin{figure}[h]
\includegraphics[bb=40 330 530 730, width=7 cm, clip]{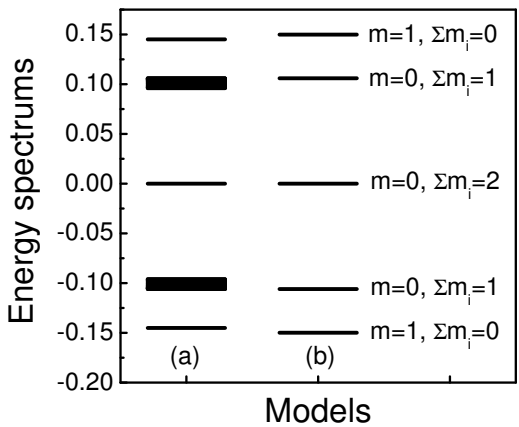}
\caption{The spectrums of $H_{s}$ (a) and $H_{c}-\protect\omega _{z}$ (b)
for $N=80$ and $I_{0}=1/2$ system in the subspace $J_{z}=-N/2+1$. The slight
deviation between the two spectrums is due to the hardcore effect of two
bosons and should vanish in macroscopic limit.}
\end{figure}

\section{Quantum Information Storage as a Dynamical Process}

We notice that the above effective Hamiltonian is just of the JC type on the
resonance and then it can be used to produce the entanglement between the
qubit state of the electron and the bosonic mode of the collective
excitation of the nuclei ensemble. Thus this entanglement induces a writing
process of qubit information into the collective excitation.

We assume that the initial state of the total system
\begin{equation}
\left\vert \psi (0)\right\rangle =\left\vert d(0)\right\rangle _{e}\otimes
\left\vert G\right\rangle
\end{equation}%
contains the arbitrary superposition%
\begin{equation}
\left\vert d(0)\right\rangle _{e}=\alpha \left\vert \uparrow \right\rangle
_{e}+\beta \left\vert \downarrow \right\rangle _{e}
\end{equation}%
of electronic states and define the Fock state%
\begin{equation}
\left\vert m\right\rangle _{b}=\frac{1}{\sqrt{m!}}(B^{+})^{m}\left\vert
G\right\rangle
\end{equation}%
for the collective excitation. We now consider the long time evolution by
projecting the wave function onto each invariant subspace spanned by the
states $\left\{ \left\vert \uparrow \right\rangle _{e}\left\vert
m\right\rangle _{b},\left\vert \downarrow \right\rangle _{e}\left\vert
m+1\right\rangle _{b}\right\} $ (we denote this subspace by $Z_{m}$). Then
this evolution of mode $B$ and the electron spin can be explicitly
characterized by reduced evolution matrices
\begin{equation}
U_{m}(t)=e^{-i\omega _{z}mt}\left[
\begin{array}{cc}
\cos \Omega _{m}t & -i\sin \Omega _{m}t \\
-i\sin \Omega _{m}t & \cos \Omega _{m}t%
\end{array}%
\right]
\end{equation}%
for $m=0,1,2,...$. Here the dressed Rabit frequency
\begin{equation}
\Omega _{m}=\sqrt{(m+1)}\Omega
\end{equation}%
depends on the number of collective excitation. A storage process of qubit
information is expressed by the factorization map
\begin{eqnarray}
\left\vert d(0)\right\rangle _{e}\otimes \left\vert G\right\rangle
&\longrightarrow &\left\vert \downarrow \right\rangle _{e}(\beta \left\vert
G_{1}\right\rangle -i\alpha e^{-i\omega _{z}\pi /2\Omega }\left\vert
G_{2}\right\rangle )  \notag \\
&=&\left\vert \downarrow \right\rangle _{e}W(\alpha \left\vert
G_{1}\right\rangle +\beta \left\vert G_{2}\right\rangle )
\end{eqnarray}%
at $t=\pi /2\Omega $. Here $\left\vert G_{1}\right\rangle =\left\vert
G\right\rangle ,\left\vert G_{2}\right\rangle =(B^{+})\left\vert
G\right\rangle $ are two collective excitation states which are orthogonal
with each other, and
\begin{equation}
W=\left[
\begin{array}{cc}
0 & 1 \\
-ie^{-i\omega _{z}\pi /2\Omega } & 0%
\end{array}%
\right]
\end{equation}%
is a unitary transformation.

The same case was also encountered in Ref. [10] (see the
difference between the initial and final states in Eq. (2) of Ref.
[10]). In quantum information theory this can be easily
implemented by a local unitary transformation independent of the
initial state $\left\vert d(0)\right\rangle _{e}$ (or the
coefficients $\alpha $ and $\beta )$ So the decoding process can
easily map back from the final state of the quantum memory
\begin{equation}
\left\vert F\right\rangle =\beta \left\vert G_{1}\right\rangle -i\alpha
e^{-i\omega _{z}\pi /2\Omega }\left\vert G_{2}\right\rangle
\end{equation}%
by its inverse transformation $W^{-1}$. In this sense, we say that the above
map implements the quantum information storage. We notice that this is very
similar to the case in quantum teleportation, in which the initial
state-independent transformation can easily be implemented by Bob for the
teleported state once the 2-bit classical information is told by Alice.

However, if one prepares the quantum memory not in its perfectly polarized
state $\left\vert G\right\rangle $ (e.g. $\left\vert m\right\rangle _{b}$ ($%
m\neq 0$)), the general initial state%
\begin{equation}
\left\vert \psi (0)\right\rangle =(\alpha \left\vert \uparrow \right\rangle
_{e}+\beta \left\vert \downarrow \right\rangle _{e})\otimes \left\vert
m\right\rangle _{b}
\end{equation}%
will evolve into%
\begin{eqnarray}
\left\vert \psi (t)\right\rangle &=&\alpha U_{m}(t)\left\vert \uparrow
\right\rangle _{e}\left\vert m\right\rangle _{b}+\beta U_{m-1}(t)\left\vert
\downarrow \right\rangle _{e}\left\vert m\right\rangle _{b}  \notag \\
&=&\alpha e^{-im\omega _{z}t}\cos \left( \Omega _{m}t\right) \left\vert
\uparrow \right\rangle _{e}\left\vert m\right\rangle _{b}  \notag \\
&&+\beta e^{-i(m-1)\omega _{z}t}\cos \left( \Omega _{m-1}t\right) \left\vert
\downarrow \right\rangle _{e}\left\vert m\right\rangle _{b}  \notag \\
&&-i\beta e^{-i(m-1)\omega _{z}t}\sin \left( \Omega _{m-1}t\right)
\left\vert \uparrow \right\rangle _{e}\left\vert m-1\right\rangle _{b}
\notag \\
&&-i\alpha e^{-im\omega _{z}t}\sin \left( \Omega _{m}t\right) \left\vert
\downarrow \right\rangle _{e}\left\vert m+1\right\rangle _{b}.
\end{eqnarray}%
It is noticed that in order to obtain the above result, we have considered $%
\left\vert \uparrow \right\rangle _{e}\otimes \left\vert m\right\rangle _{b}$
and $\left\vert \downarrow \right\rangle _{e}\otimes \left\vert
m\right\rangle _{b}$ which belong to different subspaces $Z_{m}$ and $%
Z_{m-1} $. Hence they are driven by two different blocks $U_{m}(t)$ and $%
U_{m-1}(t)$ of the block-diagonal evolution matrix $U=diag[U_{m}(t)]$,
respectively. The above result from a straightforward calculation shows that
only the ensemble of nuclei, which is prepared in the collective ground
state, the polarized ensemble, can serve as a quantum memory. Otherwise
there must exist the systematic error for quantum information processing.

\section{Decoherence due to the Couplings with Auxiliary Modes}

Finally we need to revisit the quantum decoherence in the process of quantum
information storage based on the collective excitation of the polarized
nuclei ensemble. The main source of decoherence is due to the existence of
the single particle motion described by the perturbation Hamiltonian $H_{p}$%
. The similar situation was ever considered for the collective excitation in
the ensemble of free atoms by one of the present authors (CPS) and his
collaborators \cite{sun-you}. The condition under which we can ignore the
perturbation result from non-collective excitations is just of preserving
quantum coherence. By making use of the boson modes $C_{k}$, we can expect
that the part $H_{p}$ contains the coupling between the spin qubit and $N-1$
auxiliary $C_{k}$-modes. This will realize a typical quantum decoherence
model for a two level system coupled to a bath of many harmonic oscillators.

In order to analyze this problem more quantitatively, we describe the single
particle motion from the perturbation Hamiltonian $H_{p}$ in terms of the
excitation of auxiliary modes. We consider a quasi-homogeneous case $%
g_{j}\approx g$ , in which
\begin{equation}
H_{p}\approx \sum_{k}(\omega _{z}+\frac{1}{2}g\sigma _{z}C_{k}^{+}C_{k})
\end{equation}%
where we ignore the coupling term $gB^{+}B\sigma _{z}/2\ $since it can only
lead a phase shift in the spin qubit. We consider the nuclear ensemble
prepared in a thermal equilibrium state
\begin{equation}
\rho _{R}=\frac{1}{Z}\prod_{k}\sum_{n_{k}}\exp (-\frac{\omega _{z}n_{k}}{%
k_{B}T})|n_{k}\rangle \langle n_{k}|  \notag
\end{equation}%
where
\begin{equation}
Z=\prod_{k}\sum_{n_{k}}\exp (-\frac{\omega _{z}n_{k}}{k_{B}T})
\end{equation}%
the partition function at the temperature $T$ where $k_{B}$ is the
Boltzman constant. Let the spin qubit be initially in a pure state
$\left\vert \phi \right\rangle =\alpha \left\vert 0\right\rangle
+\beta \left\vert 1\right\rangle$. After a straightforward
calculation we obtain the density matrix at time $t$
\begin{equation}
\rho \left( t\right) =U\left( t\right) (\left\vert \phi \right\rangle
\left\langle \phi \right\vert \otimes \rho _{R})U^{-1}\left( t\right)
\end{equation}%
and the corresponding reduced density matrix $\rho _{S}\left( t\right)
=Tr_{B}\rho \left( t\right) $ by tracing over the auxiliary modes $\{C_{k}\}$%
. The off-diagonal elements of $\rho _{S}\left( t\right) $ can be given
explicitly as
\begin{eqnarray}
\rho _{S}^{\star }\left( t\right) _{10} &=&\rho _{S}\left( t\right) _{01} \\
&=&\frac{\alpha ^{\star }\beta (e^{\frac{\omega _{z}}{k_{B}T}%
}-1)^{N-1}e^{i\left( N-1\right) \theta }}{\sqrt{(e^{\frac{2\omega _{z}}{%
k_{B}T}}-2e^{\frac{\omega _{z}}{k_{B}T}}\cos (gt)+1)^{N-1}}}  \notag
\end{eqnarray}%
where
\begin{equation}
\theta =\arctan \frac{\sin \left( gt\right) }{\exp \left( \frac{\omega _{z}}{%
k_{B}T}\right) -\cos \left( gt\right) }
\end{equation}%
In the zero temperature limit or $T\rightarrow 0$, there is no decoherence
since the off-diagonal elements
\begin{equation}
\rho _{S}^{\star }\left( t\right) _{10}=\rho _{S}\left( t\right)
_{01}=\alpha ^{\star }\beta
\end{equation}%
do not change. But in a finite temperature, the norm of off-diagonal element
is proportional to the so-called decoherence factor
\begin{eqnarray}
D(T,t) &=&d(T,t)^{N-1}  \notag \\
&\equiv &\frac{(e^{\frac{\omega _{z}}{k_{B}T}}-1)^{N-1}}{\sqrt{(e^{\frac{%
2\omega _{z}}{k_{B}T}}-2e^{\frac{\omega _{z}}{k_{B}T}}\cos (gt)+1)^{N-1}}}.
\end{eqnarray}%
This result illustrates that thermal excitation of the auxiliary modes will
block the implementation of the macroscopic nuclear ensemble based quantum
memory since in the large $N$ limit $D(T,t)\rightarrow 0$ except for the
special instances at
\begin{equation}
t=\frac{2k\pi }{g},k=0,1,2...
\end{equation}%
In these instances, $D(T,t)=1$ and there is no decoherence at all. Besides,
since in these instances $\rho _{S}^{\star }\left( t\right) _{10}=\alpha
^{\star }\beta $ is just the initial values and then we implement a ideal
quantum information storage to recover the stored state. To further consider
the temperature dependecne of the auxiliary mode induced decoherence, we
plot a $3D$-graphic of $D(T,t)$ for a small size $N=20$ system. It shows
that the case $D(T,t)=1$ appears periodically as time $t$, and $%
D(T,t)\rightarrow 1$ all the time when $T\rightarrow 0$. According to the
experimental data \cite{uqm-zoller}, the period is roughly estimated as $%
2\pi /g\sim 10^{-7}s$.
\begin{figure}[h]
\includegraphics[bb=42 206 530 611, width=7 cm, clip]{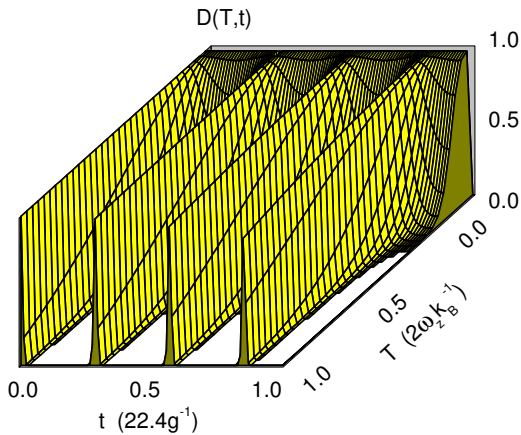}
\caption{The decoherence factor $D(T,t)$ for a small size system as the
function of time and temperature.}
\end{figure}

\section{Summary with Remarks}

In summary we have studied the possibility of quantum memory by using
collective excitation of ensemble of polarized nuclei surrounding a single
electronic spin in a quantum dot. We explicitly present the
quasi-homogeneous independent conditions, under which the many-particle
system, a macroscopic ensemble of polarized nuclei, can be treated as a
single mode bosonic system. Thus the interaction is of the similar form of
Jaynes-Cummings model. Based on this fact, the collective excitation can
serve as a quantum memory to store the spin state of a conduction electron.

We also pointed out that the physical system for quantum
information storage is the same as that in Ref. [10] which first
showed that electronic spin coherence can be reversibly mapped
onto the collective state of the surrounding nuclei. But our
studies emphasize that the collective excitation-based quantum
memory can be understood in terms of the spin-boson model with
essential simplicity in physics. Especially the valid conditions
are discovered in present paper. That is, the collective operators
are explicitly invoked to depict the bosonic collective
excitations and then we can present an effective boson-spin model,
which reveals physical mechanism with collective quantum coherence
behind the original conceptual protocol for the long-lived quantum
memory.

There are two sources of quantum decoherence in such quantum
information processing, one is due to non-collective mode and the
other is due to the nuclear spin diffusion or coupling with
environment. The latter is dominate and has been well considered
in Ref. \cite{Phi}, but the former can still play a role in
certain cases. So we stress the former in this paper since the
same situation was even considered for the collective excitation
in the ensemble of free atoms by us \cite{sun-you}. In principle,
the latter can also be treated in our spin-boson model with
similar approach by adding diffusion terms. We also noticed that
the systematic errors in transferring quantum information can
occur due to the appearance of higher excitation by illustrating
that only the ensemble of nuclei prepared in the collective ground
state rather than the excited ones can serve as a quantum memory.
How to avoid the higher excitation of the collective boson mode
and how to correct the error due to the appearance of higher
excitation are open questions that need further investigations.

We acknowledge the support of the CNSF (grant No. 90203018, 10474104), the
Knowledge Innovation Program (KIP) of the Chinese Academy of Sciences, the
National Fundamental Research Program of China (No. 001GB309310) and Science
and Technology Cooperation Fund of Nankai and Tianjin University.

\end{document}